\documentclass[twocolumn]{aastex6}

\usepackage{natbib}

\newcommand{\kms}{km\,s$^{-1}$}
\newcommand{\cii}{[\ion{C}{2}]}
\newcommand{\ci}{[\ion{C}{1}]}
\newcommand{\lfir}{$L_{\mathrm{FIR}}$}
\newcommand{\ltir}{$L_{\mathrm{TIR}}$}
\newcommand{\lcii}{$L_\mathrm{[CII]}$}
\newcommand{\lci}{$L_\mathrm{[CI]}$}
\newcommand{\lsun}{$L_\sun$}
\newcommand{\msun}{$M_\sun$}
\newcommand{\msunyr}{$M_\sun$\,yr$^{-1}$}

\newcommand{\mbh}{$M_\mathrm{BH}$}

\slugcomment{Accepted for publication in ApJ}

\shorttitle{The compact, $\sim$1\,kpc host galaxy of a quasar at $z=7.1$}
\shortauthors{Venemans et al.}

\begin{document}

\title{The compact, $\sim$1\,kpc host galaxy of a quasar at a redshift of 7.1}

\author{Bram P. Venemans\altaffilmark{1},
Fabian Walter\altaffilmark{1,2,3},
Roberto Decarli\altaffilmark{1},
Eduardo Ba\~nados\altaffilmark{4,12},
Jacqueline Hodge\altaffilmark{5},
Paul Hewett\altaffilmark{6},
Richard G. McMahon\altaffilmark{6,7},
Daniel J. Mortlock\altaffilmark{8,9,10},
Chris Simpson\altaffilmark{11}
}
\altaffiltext{1}{Max-Planck Institute for Astronomy, K{\"o}nigstuhl 17, 69117 Heidelberg, Germany}
\altaffiltext{2}{Astronomy Department, California Institute of Technology, MC105-24, Pasadena, CA 91125, USA}
\altaffiltext{3}{National Radio Astronomy Observatory, Pete V. Domenici Array Science Center, P.O. Box O, Socorro, NM 87801, USA}
\altaffiltext{4}{The Observatories of the Carnegie Institute of Washington, 813 Santa Barbara Street, Pasadena, CA 91101, USA}
\altaffiltext{5}{Leiden Observatory, Leiden University, P.O. Box 9513, NL2300 RA Leiden, The Netherlands}
\altaffiltext{6}{Institute of Astronomy, University of Cambridge, Madingley Road, Cambridge CB3 0HA, UK}
\altaffiltext{7}{Kavli Institute for Cosmology, University of Cambridge, Madingley Road, Cambridge CB3 0HA, UK}
\altaffiltext{8}{Astrophysics Group, Blackett Laboratory, Imperial College London, London, SW7 2AZ, UK}
\altaffiltext{9}{Department of Mathematics, Imperial College London, London, SW7 2AZ, UK}
\altaffiltext{10}{Department of Astronomy, Stockholm University, Albanova, SE-10691 Stockholm, Sweden}
\altaffiltext{11}{Gemini Observatory, 670 N. A'ohoku Place, Hilo, Hawaii, 96720, USA}
\altaffiltext{12}{Carnegie-Princeton Fellow}

\begin{abstract}
We present ALMA observations of the \cii\ fine structure line and the
underlying far--infrared (FIR) dust continuum emission in J1120+0641,
the most distant quasar currently known ($z\!=\!7.1$). We also present
observations targeting the CO(2-1), CO(7-6) and \ci\ 369\,$\mu$m lines
in the same source obtained at the VLA and PdBI.  We find a \cii\ line
flux of $F_\mathrm{[CII]}=1.11\pm0.10$\,Jy\,\kms\ and a continuum flux
density of $S_\mathrm{227\,GHz}=0.53\pm0.04$\,mJy\,beam$^{-1}$,
consistent with previous unresolved measurements.  No other source is
detected in continuum or \cii\ emission in the field covered by ALMA
($\sim$\,25\arcsec).  At the resolution of our ALMA observations
(0\farcs23, or 1.2\,kpc, a factor $\sim$70 smaller beam area compared
to previous measurements), we find that the majority of the emission
is very compact: a high fraction ($\sim$80\%) of the total line and
continuum flux is associated with a region 1--1.5\,kpc in
diameter. The remaining $\sim$20\% of the emission is distributed over
a larger area with radius $\lesssim$4\,kpc. The \cii\ emission does
not exhibit ordered motion on kpc-scales: applying the virial theorem
yields an upper limit on the dynamical mass of the host galaxy of
$(4.3\pm0.9)\times10^{10}$\,\msun, only $\sim$20$\times$ higher than
the central black hole. The other targeted lines (CO(2-1), CO(7-6) and
\ci) are not detected, but the limits of the line ratios with respect
to the \cii\ emission imply that the heating in the quasar host is
dominated by star formation, and not by the accreting black hole.  The
star-formation rate implied by the FIR continuum is 105--340\,\msunyr,
with a resulting star-formation rate surface density of
$\sim$100--350\,\msunyr\,kpc$^{-2}$, well below the value for
Eddington--accretion--limited star formation.
\end{abstract}


\section{INTRODUCTION}

Luminous quasars are powered by accreting, supermassive black holes
(BHs). Quasars and thus supermassive BHs have been found at high
redshifts, $z\sim7$, when the Universe was less than a billion year
old \citep[e.g.,][]{mor11,ven13}. These early, supermassive black
holes are generally hosted by massive galaxies that form stars at a
high rate. Indeed, locally there is a relation between the mass of the
bulge and the mass of the black hole in its center \citep[see,
  e.g.,][for a review]{kor13}. An important question in astronomy is
how the first black holes formed and whether the black hole and
hosting galaxy coevolved.

Over the last 15 years, numerous surveys established a sample of
$\sim$100 quasars at $z>6$
\citep[e.g.,][]{fan06b,mor09,jia09,wil10a,ven13,carn15,jia15,ree15,ven15a,ban16,mat16}. The
most luminous of these have BHs with masses in excess
$10^9$\,\msun\ \citep[e.g.,][]{jia07,kur07,wil10b,der11,der14,wu15,ven15a}. As
the accreting black hole generally dominates the emission at
rest-frame UV and optical wavelengths, observations at
(sub-)millimeter wavelength are needed to study the galaxies hosting
these black holes. Several groups have targeted and detected
$z\gtrsim6$ quasars with mm facilities such as the IRAM Plateau de
Bure Interferometer (PdBI) and Atacama Large Millimeter/submillimeter
Array (ALMA)
\citep[e.g.,][]{ber03a,mai05,wan08b,wan13,wil13,wil15,ven16}. These
data show that rapid black hole growth is, in some cases, accompanied
by very high star-formation rates (SFR) of up to $\sim$1000\,\msunyr.

In this paper we investigate the host galaxy of the most distant
quasar currently known, ULAS J112001.48+064124.3 (hereafter,
J1120+0641) at a redshift of $z=7.085$ \citep{mor11}. The quasar is
powered by a black hole with a mass of
$(2.4\pm0.2)\times10^9$\,\msun\ \citep{mor11,der14} and is accreting
close to the Eddington limit \citep{mor11,der14,bar15}. The quasar
host galaxy has previously been detected with the 
Interferometer IRAM PdBI in \cii\ and the dust continuum
\citep{ven12}. In these data the host galaxy was unresolved in a
$\sim$2\arcsec\ beam and the dynamical mass and the morphology of the
line emitting gas could not be constrained. Here we present high
spatial resolution (0\farcs23, or 1.2\,kpc) observations with ALMA
(Section~\ref{sec:almaobs}), decreasing the beam area by a factor of
$\sim$70. We also present observations with the PdBI
(Section~\ref{sec:pdbiobs}) and the NRAO Karl G. Jansky Very Large
Array (VLA, Section~\ref{sec:evlaobs}) targeting the neutral and
molecular gas lines CO(7-6), \ci\ 369\,$\mu$m, and CO(2-1).  Our
results are detailed in Section~\ref{sec:results}, followed by a
discussion about the implications of our findings in
Section~\ref{sec:discussion}. A summary is presented in
Section~\ref{sec:summary}.

Throughout this paper, we adopt a concordance cosmology with $H_0=70$
km\,s$^{-1}$\,Mpc$^{-1}$, $\Omega_M=0.3$, and $\Omega_\lambda=0.7$,
leading to a spatial scale of 5.2\,proper kpc per arcsec at
$z=7.085$. Star formation rates (SFRs) are calculated assuming a
\citet{kro01} initial mass function (IMF).

\section{OBSERVATIONS}
\label{sec:observations}

\subsection{ALMA Cycle 1 Observations}
\label{sec:almaobs}

The host galaxy of J1120+0641 was observed with ALMA between 2014 June
and 2015 June (program 2012.1.00882.S). In 2014 the antennas were in a
compact configuration (33--36 antennas with baselines between
20--650\,m) and in 2015 in a more extended configuration, with
baselines between 34--1574\,m and a total of 38--47 antennas. On 2014
June 9, J1120+0641 was observed for 33.5\,min (on source), on 2014
June 10 for 26.8\,min, and on 2014 June 11 for 33.5\,min. The combined
2014 data reached an rms noise of 0.39\,mJy\,beam$^{-1}$ in a 20\,MHz
bin and the beam size was 0\farcs62$\times$0\farcs51 (natural
weighting). On 2015 June 26 and 27, the quasar was observed for
33.4\,min (on source) during both days. The beam size of the combined
2015 data was 0\farcs25$\times$0\farcs24 (using natural weighting) and
the rms noise per 20\,MHz bin is 0.21\,mJy\,beam$^{-1}$. The full
dataset reached a noise of 0.15\,mJy\,beam$^{-1}$\,(20\,MHz)$^{-1}$
and has a beam size of 0\farcs31$\times$0\farcs29 using natural
weighting. Using a weighting factor of robust\,=\,0.5 results in a
slightly higher rms noise of 0.17\,mJy\,beam$^{-1}$ per 20\,MHz bin
and a beamsize of 0\farcs23$\times$0\farcs22. This corresponds to
1.2\,kpc at the redshift of the quasar.

The ALMA observations covered the redshifted \cii\ line at
$\nu_\mathrm{obs}=235.07$\,GHz with two overlapping bandpasses of
1.875\,GHz each. The overlap was 20\% which resulted in frequency
coverage of 3.375\,GHz around the \cii\ line. Two additional
bandpasses of 1.875\,GHz each were placed around an observed frequency
of 220\,GHz. The data were reduced using Common Astronomy Software
Applications \citep[CASA,][]{mul07}, following standard reduction
steps.

\subsection{PdBI 3mm Observations}
\label{sec:pdbiobs}

The CO(7-6) ($\nu_\mathrm{rest}=806.652$\,GHz) and
\ci\ ($\nu_\mathrm{rest}=809.344$\,GHz) emission lines from J1120+0641
(redshifted to observed frequency around
$\nu_\mathrm{obs}\sim100$\,GHz) were targeted by the PdBI between 2011
July 4 and 2012 May 3. The observations were carried out with 5--6
antennas. The antennas were in the most compact (D) configuration,
providing a beam with a size of 4\farcs7$\times$4\farcs2. The WideX
correlator that was used provided a continuous frequency coverage of
3.6\,GHz and covered both CO(7-6) and \ci\ emission lines in a single
setup. The data were reduced using the Grenoble Image and Line Data
Analysis System (GILDAS) software package. The total time on source
was 18.3\,hr (six antenna equivalent), resulting in an rms noise of
0.33\,mJy\,beam$^{-1}$ per 20\,MHz bin. he continuum rms noise,
measured in an image that was created by averaging all channels that
do not cover the emission lines (resulting in a continuum bandwidth of
2.7\,GHz, see Section~\ref{sec:cocilimits}), is 29\,$\mu$Jy.

\begin{figure*}
\figurenum{1}
\plottwo{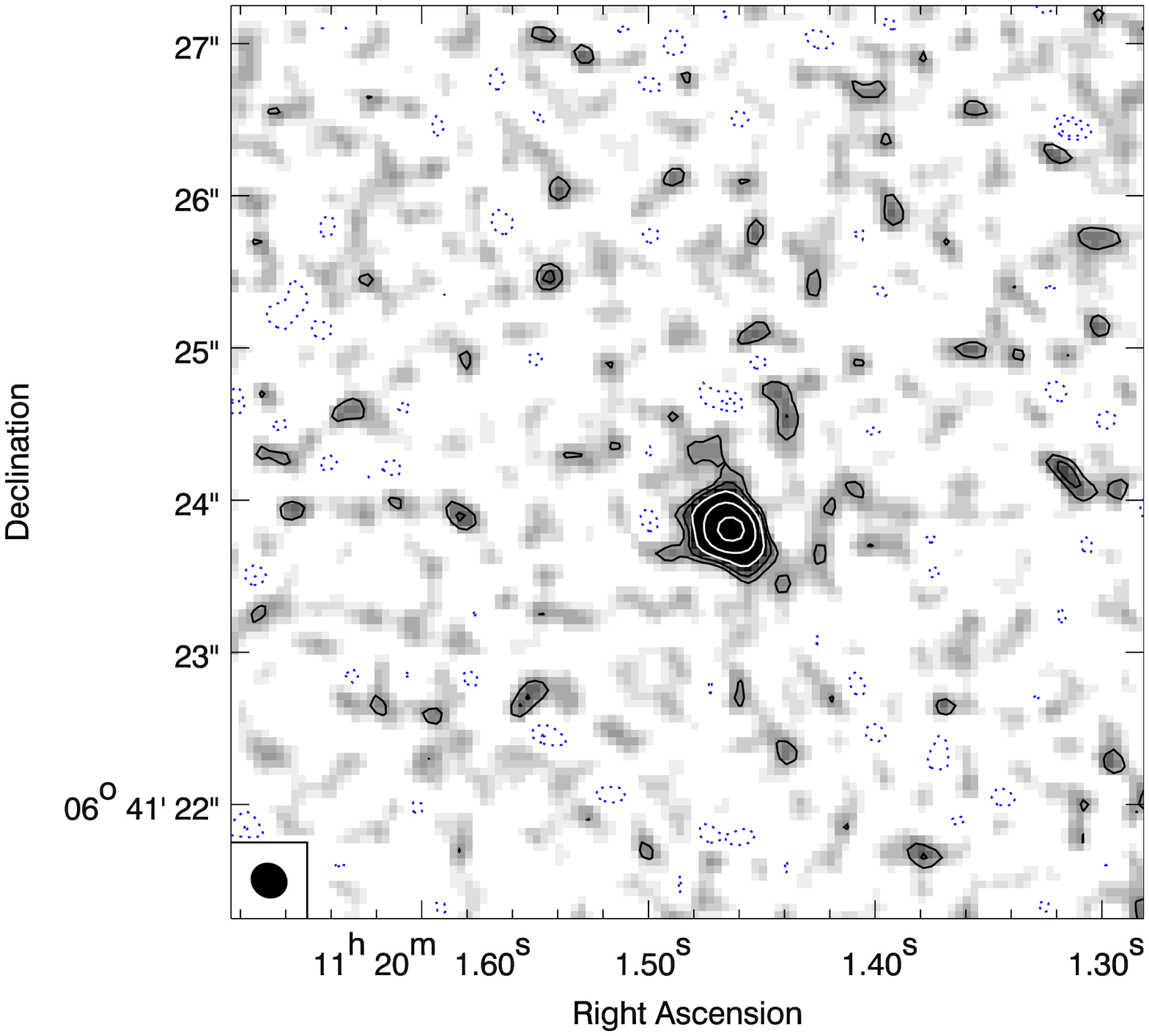}{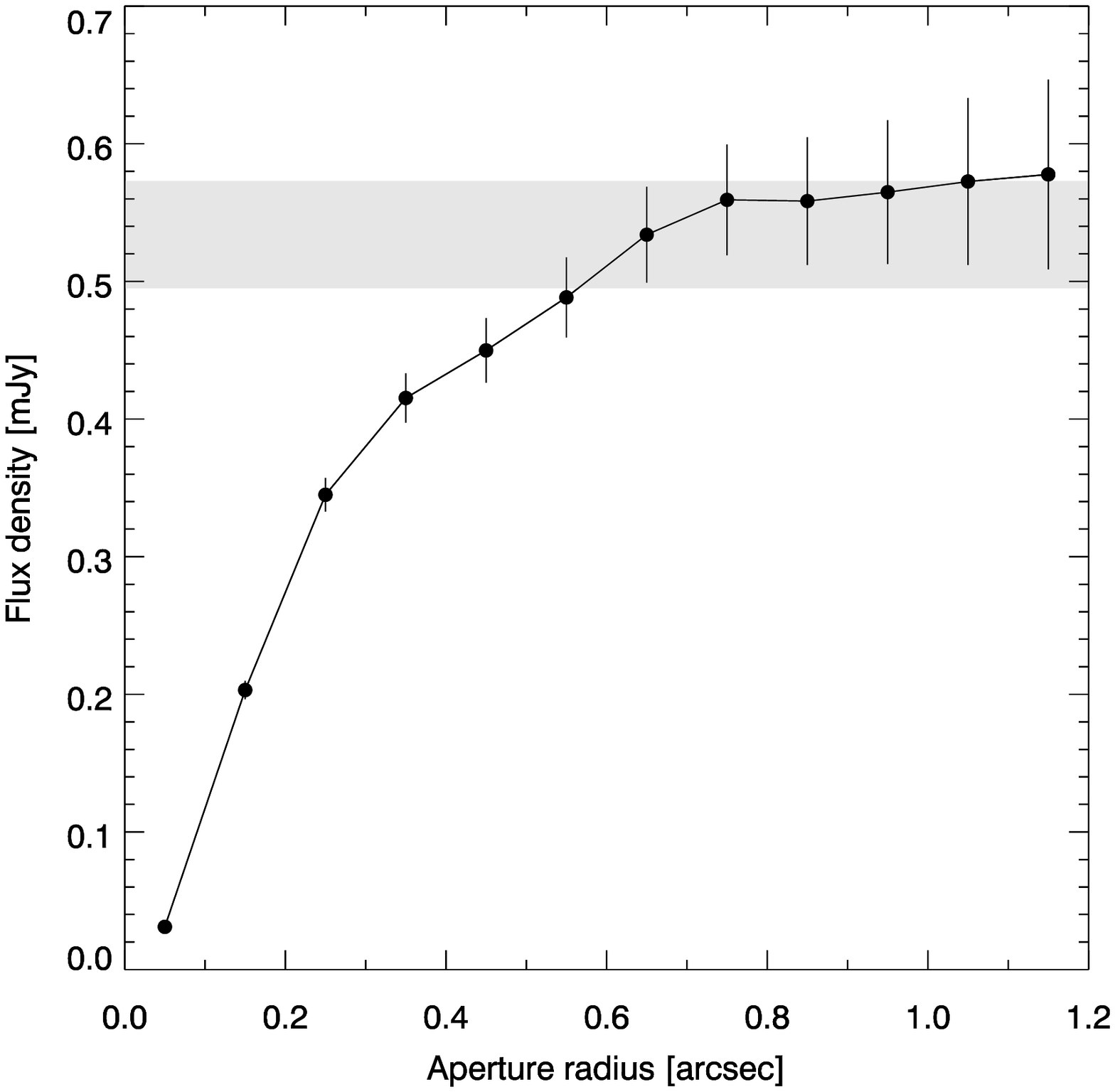}
\caption{{\it Left:} 227\,GHz ($\lambda_\mathrm{obs}=1.3$\,mm)
  continuum map of the field of J1120+0641. To create the map all
  channels at least 0.75\,GHz away from the position of the
  \cii\ emission line ($\nu_\mathrm{[CII],obs}\approx235$\,GHz, see
  Figure~\ref{fig:spectrum}) were averaged. The beam of
  0\farcs25$\times$0\farcs23 is shown in the bottom-left corner. The
  1$\sigma$ rms noise in this map is 11\,$\mu$Jy\,beam$^{-1}$. The
  blue, dashed contours are --3$\sigma$ and --2$\sigma$; the black,
  solid contours are +2$\sigma$ and +3$\sigma$; the white solid
  contours are +5$\sigma$, +10$\sigma$, and +20$\sigma$. Only the
  central quasar is detected (S/N$\sim$26). {\it Right:} Estimated
  flux density of the central continuum source as function of aperture
  radius. The flux density reaches a roughly constant value for
  apertures with radius $>0.7$\,arcsec and the value is consistent
  within the errors with the peak flux density when observed with a
  beam of 1\farcs9 (the gray band).}
\label{fig:continuum}
\end{figure*}

\subsection{VLA Observations}
\label{sec:evlaobs}

We observed CO(2--1) in J1120+0641 (rest frequency $\nu_{rest}$ =
230.5424\,GHz, redshifted to $\nu_{obs}$ = 28.5145\,GHz) as part of
VLA project 11A-285. The observations were taken on 2013 January 12
and 13 in the DnC configuration (consisting of 27 antennas). We
utilized the Ka band receivers with the A/C IF pair tuned to
32.008\,GHz and the B/D IF pair tuned to 27.810\,GHz. The observations
were taken in full polarization mode with 16 128--MHz spectral windows
(8 spectral windows per IF) and 64 2--MHz channels per spectral
window.

The observations consisted of a single pointing centered on the quasar
host galaxy. We used fast switching phase calibration \citep{car99} on
VLA calibrator J1058$+$0133 with a four-minute cycle time, and the
same source served as our bandpass calibrator. The bright source 3C286
served as the flux calibrator for all observations. The observing
time, including overhead, was 10\,hr. The total time on source was
6.3\,hr.

The data were reduced using the CASA package. We imaged the
calibrated data using the clean algorithm with a cell size of 0\farcs5
and natural weighting, resulting in a synthesized beam of
2\farcs3$\times$1\farcs3. The final cube has an rms of
15\,$\mu$Jy\,beam$^{-1}$ in a 76\,MHz window (800\,\kms) centered on
the frequency of the expected CO line. The continuum rms is
4\,$\mu$Jy\,beam$^{-1}$.

\section{RESULTS}
\label{sec:results}

\subsection{FIR Continuum}
\label{sec:fircontinuum}

In Figure~\ref{fig:continuum} we show a 227\,GHz (observed wavelength
$\lambda_\mathrm{obs}=1.3$\,mm) continuum map of the field that was
created by averaging all channels at least 0.75\,GHz away from the
expected \cii\ emission line. As a consequence, the reconstructed
continuum emission is dominated by ALMA's lower--sideband observations
(at frequencies $\sim$\,10\,GHz below the redshifted \cii\ line). In
the whole field (with a half power beam width of 24\farcs8) only the
quasar is detected (at signal-to-noise S/N\,$>$\,5, or
$S_\mathrm{227\,GHz}>55 \mu$Jy\,beam$^{-1}$).  The position
(RA\,=\,11$^h$20$^m$01.465$^s$;
Dec\,=\,+06$^\circ$41$^\prime$23.810\arcsec) is approximately 0\farcs5
to the South of the position from the UK Infrared Telescope (UKIRT)
Infrared Deep Sky Survey (UKIDSS) published in \citet{mor11}.  This
difference could be due to a systematic difference in the absolute
astrometric calibration between UKIDSS and ALMA. It is not possible to
verify this, as there are no other sources in the field within the
ALMA FOV beam, but similar offsets between ALMA and
optical/near-infrared images have been reported (e.g., in the Hubble
Ultra-Deep Field, see \citealt{ara16,ruj16}). No offset from the
UKIDSS position was seen in the PdBI data \citep{ven12}. Within the
uncertainties of the PdBI observations, the earlier mm continuum
position was identical to that of the UKIDSS position.

\begin{table}[!t]
\caption{Observed Properties of J1120+0641 \label{tab:res}}
\begin{tabular}{lc}
\hline
\hline
RA (J2000) & 11$^h$20$^m$01.465$^s$ \\
Dec (J2000) & +06$^\circ$41$^\prime$23.810\arcsec\ \\
$z_\mathrm{[CII]}$ & 7.0851$\pm$0.0005 \\
$F_\mathrm{[CII]}$ [Jy\,\kms] & 1.11$\pm$0.10 \\
FWHM$_\mathrm{[CII]}$ [\kms] & 400$\pm$45~ \\
$S_{227\,\mathrm{GHz}}$ [mJy] & 0.53$\pm$0.04 \\
$S_{100\,\mathrm{GHz}}$ [mJy] & 0.086$\pm$0.029 \\
$S_{30\,\mathrm{GHz}}$ [mJy] & $<$0.011 \\
EW$_\mathrm{[CII]}$ [$\mu$m] & 0.90$\pm$0.14 \\
$F_\mathrm{CO(2-1)}$ [Jy\,\kms] & $<$0.034 \\
$F_\mathrm{CO(7-6)}$ [Jy\,\kms] & $<$0.20 \\
$F_\mathrm{[CI]}$ [Jy\,\kms] & $<$0.20 \\
size continuum\tablenotemark{a} [arcsec$^2$] & (0.23$\pm$0.03)$\times$(0.16$\pm$0.03) \\
size continuum\tablenotemark{a} [kpc$^2$] & (1.24$\pm$0.14)$\times$(0.83$\pm$0.14) \\
size \cii\ emission\tablenotemark{a} [arcsec$^2$] & (0.31$\pm$0.05)$\times$(0.27$\pm$0.05) \\
size \cii\ emission\tablenotemark{a} [kpc$^2$] & (1.65$\pm$0.29)$\times$(1.44$\pm$0.26) \\
\hline
\end{tabular}
\tablenotetext{a}{The size listed here are diameters and are derived for the central component that contains around 80\% of the total flux density (see Section~\ref{sec:size}).}
\end{table}

The peak flux density of the quasar host is
$S_\mathrm{227\,GHz}=0.26\pm0.01$\,mJy\,beam$^{-1}$. To measure the
total continuum flux density we tapered the emission to 1\farcs9. This
resolution is similar to the beam of the original \cii\ observations
\citep[2\farcs0$\times$1\farcs7;][]{ven12}.  In this map, the peak
flux density of the host is
$S_\mathrm{227\,GHz}=0.53\pm0.04$\,mJy\,beam$^{-1}$. A similar value
was derived when performing aperture photometry (see
Figure~\ref{fig:continuum}). For apertures larger than $\sim$0\farcs7
the flux density is roughly constant and we measure a total flux
density of $S_\mathrm{227\,GHz}=0.56\pm0.04$\,mJy (see
Section~\ref{sec:size}). Note that due to the shape of the FIR
continuum, the continuum flux density is higher around the \cii\ line
(Fig.~\ref{fig:spectrum}, see also, e.g., Section~\ref{sec:sfrd} and
\citealt{ven16}). From the spectrum of the \cii\ line (around an
observed frequency of 235\,GHz) (Figure~\ref{fig:spectrum}) we measure
$S_\mathrm{235\,GHz}=0.64\pm0.08$\,mJy\,beam$^{-1}$, which is
consistent with the published value of
$S_\mathrm{235\,GHz}=0.61\pm0.16$\,mJy\,beam$^{-1}$ \citep{ven12}.

We also created a map of the 100\,GHz
($\lambda_\mathrm{obs}\approx3$\,mm) continuum emission from the PdBI
data, using the channels that are expected to be line--free.  At the
position of the 227\,GHz continuum source we obtain a tentative
3$\sigma$ detection (flux density of
$S_{100\,\mathrm{GHz}}=86\pm29$\,$\mu$Jy). Based on the 1\,mm
continuum detection in the ALMA data, assuming an intrinsic dust
temperature of 30--50\,K (see Sections~\ref{sec:pdr} and
\ref{sec:sfrd}), an emissivity index of $\beta=1.6$ and taking the
cosmic microwave background (CMB) into account \citep[see,
  e.g.,][Section~\ref{sec:pdr}]{dac13,ven16}, the expected flux
density at 100\,GHz is 42--55\,$\mu$Jy, consistent with our low
signal-to-noise measurement. The non-detection of the continuum in the
VLA data ($S_{30\,\mathrm{GHz}}<11$\,$\mu$Jy, Table~\ref{tab:res}) is
consistent with J1120+0641 being radio-quiet \citep{mom14}.

\subsection{\cii\ Emission Line}
\label{sec:ciiemission}

\begin{figure}
\figurenum{2}
\includegraphics[width=\columnwidth]{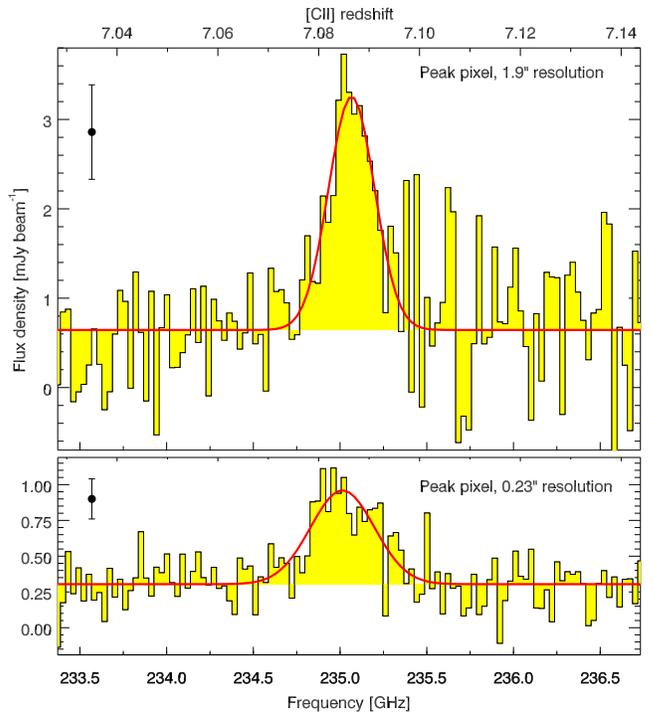}
\caption{{\it Top:} Spectrum extracted from the peak pixel in the ALMA
  data cube tapered to a spatial resolution of 1\farcs9. The bins are
  30\,MHz, which corresponds to $\sim$38\,\kms. The typical 1$\sigma$
  uncertainty per bin of 0.52\,mJy\,beam$^{-1}$ is plotted in the
  top-left corner. The red line is a Gaussian+constant fit to the
  spectrum. {\it Bottom:} Same as above, but this spectrum is
  extracted from the peak pixel in the data cube with the full spatial
  resolution (0\farcs23). The rms noise is 0.14\,mJy\,beam$^{-1}$ per
  30\,MHz bin.\vspace{0.5cm}}
\label{fig:spectrum}
\end{figure}

The \cii\ emission line is detected at high S/N in the ALMA data. To
get an estimate of the total line flux we tapered the data cube to a
beam of 1\farcs9.  We show the spectrum from the peak pixel in this
cube, together with a Gaussian fit, in Figure~\ref{fig:spectrum}
(top). The redshift of the \cii\ line is
$z_{\mathrm{[CII]}}=7.0851\pm0.0005$, the peak flux density is
$f_p=2.60\pm0.25$\,mJy\,beam$^{-1}$, and the full width at half
maximum is FWHM$_\mathrm{[CII]}=400\pm45$\,\kms\ (see also
Table~\ref{tab:res}). The line flux of
$F_{\mathrm{[CII]}}=1.11\pm0.10$\,Jy\,\kms\ is consistent with the
value of $F_{\mathrm{[CII]}}=1.03\pm0.14$\,Jy\,\kms\ published by
\citet{ven12}, while the line width measured in the ALMA data is a
factor $1.69\pm0.32$ larger than the earlier value that was based on
lower S/N data. The \cii\ rest-frame equivalent width is
EW$_\mathrm{[CII]}=0.90\pm0.14$\,$\mu$m. This is only $\sim$30\% lower
than that of local starburst galaxies \citep[e.g.,][]{dia13,sar14}.

\begin{figure*}
\figurenum{3}
\plottwo{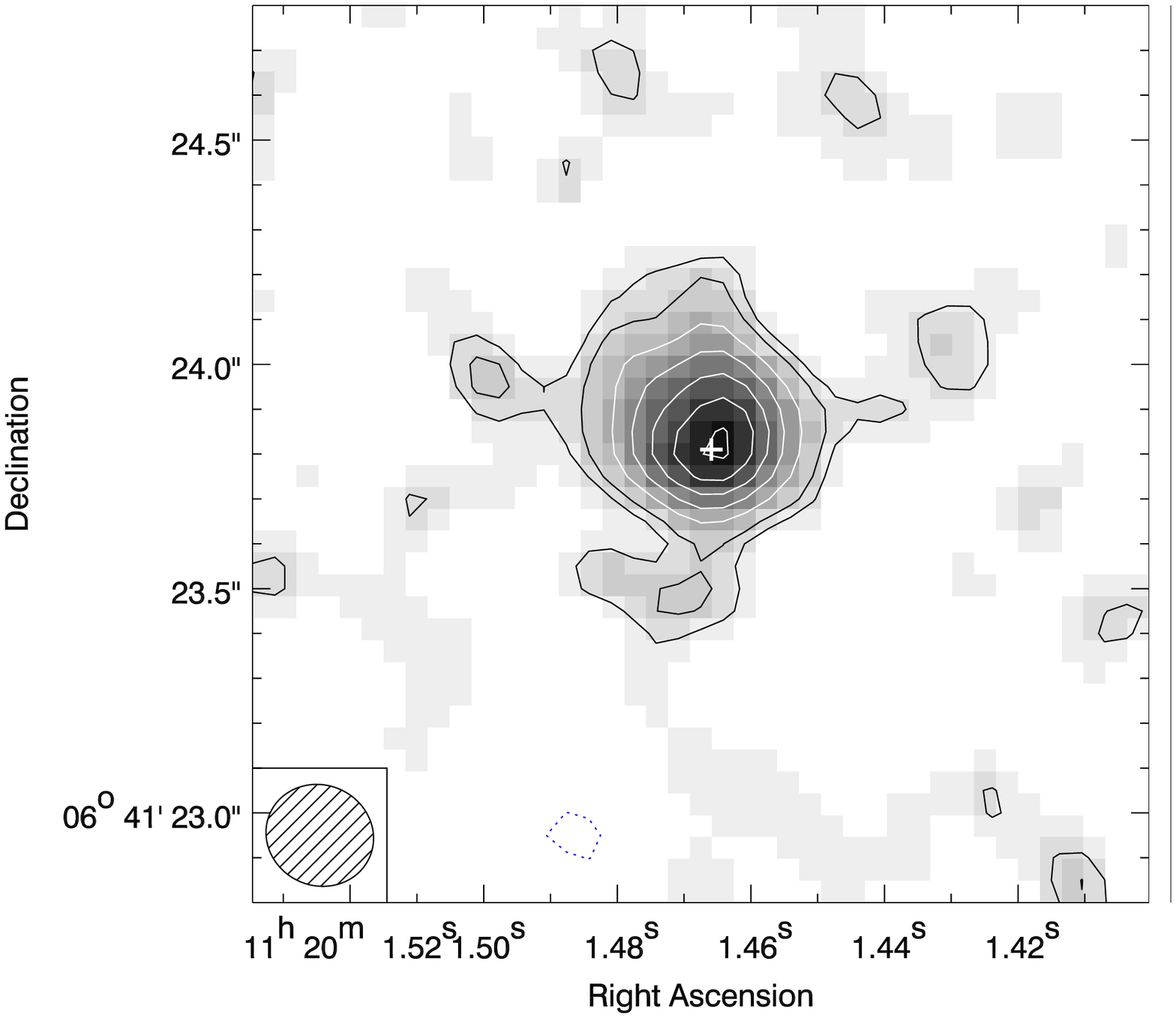}{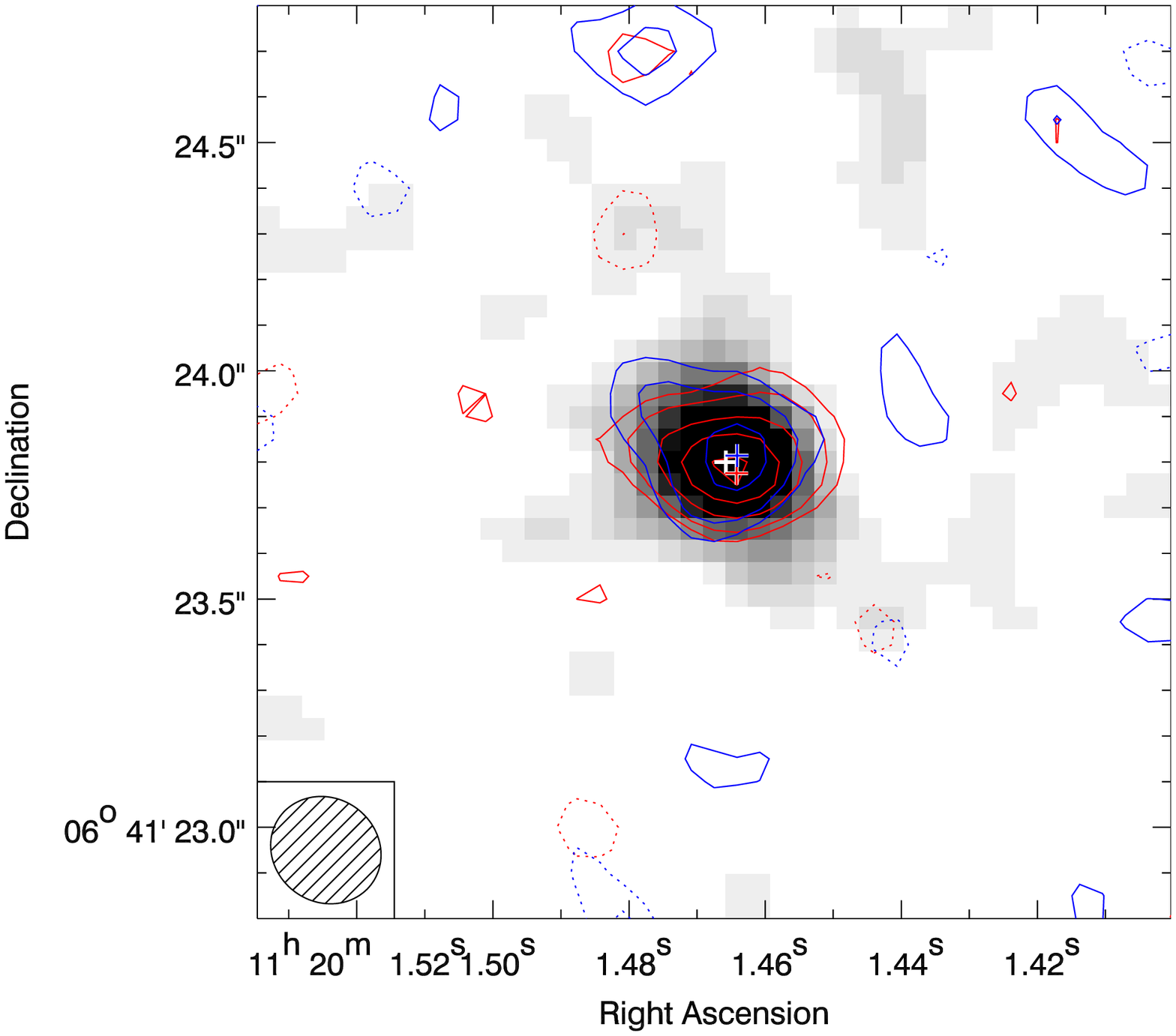}
\caption{{\it Left:} Map of the \cii\ emission in J1120+0641 (shown
  both in greyscale as well as contours), averaged over
  $1.5\times$\,FWHM$_\mathrm{[CII]}$ (600\,\kms\ or 470\,MHz). The
  1$\sigma$ rms noise in this map is 42\,$\mu$Jy. The blue, dashed
  contours are --3$\sigma$ and --2$\sigma$; the black, solid contours
  are +2$\sigma$ and +3$\sigma$; the white solid contours are
  [5,7,9,11,13]\,$\times\sigma$. The size of the beam is shown in the
  bottom-left corner. The white cross shows where the continuum
  emission of the quasar host peaks. {\it Right:} The red and blue
  side of the emission line are shown in contours, averaged over
  265\,\kms\ centered at +265\,\kms\ (red contours) and
  --265\,\kms\ (blue contours) from the line peak. Contour levels are
  --2$\sigma$, +2$\sigma$, +3$\sigma$, +5$\sigma$, +7$\sigma$, and
  +9$\sigma$, with $\sigma\sim$57\,$\mu$Jy. The greyscale is a
  representation of the continuum map. The white, red, and blue
  crosses indicate the peak of the total, redshifted, and blueshifted
  \cii\ emission, respectively. }
\label{fig:ciimaps}
\end{figure*}

Also shown in Figure~\ref{fig:spectrum} (bottom) is the spectrum of
the peak pixel in the high resolution data cube (with a beam of
0\farcs23$\times$0\farcs22). The \cii\ line is significantly fainter
in the centre with a peak flux density
$f_p=0.66\pm0.06$\,mJy\,beam$^{-1}$ and a flux of
$F_{\mathrm{[CII]}}=0.39\pm0.04$\,Jy\,\kms, which is 35\% of the total
line flux. This means that the \cii\ line is spatially resolved in our
data (see Section~\ref{sec:size}). The peak of the \cii\ emission
coincides with the peak of the continuum emission
(Figure~\ref{fig:ciimaps}). The line width is broader in the centre
with a FWHM$_\mathrm{[CII]}=555\pm60$\,\kms. We will come back to this
in Section~\ref{sec:mdyn}.

\subsection{Limits on CO and \ci\ Emission}
\label{sec:cocilimits}

We searched for CO(7-6) and \ci\ emission in the PdBI 3\,mm
data. After creating a continuum subtracted data cube using the CASA
task ``uvcontsub", we averaged the channels where we expected the
CO(7-6) and \ci\ lines, based on the \cii\ redshift, over
400\,\kms\ (the \cii\ line FWHM, see Table~\ref{tab:res}). In the
resulting maps (not shown here) no significant ($>$3$\sigma$) line
emission was detected at the location of the quasar host galaxy. For
the CO(7-6) and \ci\ lines we derived 3$\sigma$ upper limits on the
line strength of
$F_\mathrm{CO(7-6),[CI]}<0.20$\,Jy\,\kms\footnote{These upper limits
  are corrected to account for flux in the outer linewings that are
  not included when averaging over the FWHM of a Gaussian line.}.

Similarly, we derived an upper limit on the CO(2-1) line from the VLA
data. Averaging the data over 400\,\kms\ resulted in a map with no
significant emission at the position of the quasar host. The rms of
the image was 0.021\,mJy and the 3$\sigma$ upper limit on the CO(2-1)
line was $F_\mathrm{CO(2-1)}<0.034$\,Jy\,\kms. We will discuss the
implications of these non-detections in Section~\ref{sec:pdr}.

\subsection{Size and Structure of the Emission Regions}
\label{sec:size}

{\em Continuum:} To estimate the size of the continuum emitting
region, we fitted a 2D Gaussian to the continuum map using the CASA
task ``imfit". The quasar host galaxy is marginally resolved and we
derive a deconvolved size with a FWHM of
(0.23$\pm$0.03)$\times$(0.16$\pm$0.03)\,arcsec$^2$, or
(1.24$\pm$0.14)$\times$(0.83$\pm$0.14)\,kpc$^2$. The integrated flux
density of this central component is $0.43\pm0.03$\,mJy, which is
$\sim$80\% of the peak measured in the tapered continuum image
(Section~\ref{sec:fircontinuum}). The remaining 20\% of the total
continuum flux density comes from a more extended region, with a size
$\lesssim$0\farcs6 {in radius ($\lesssim$3\,kpc, see
  Figure~\ref{fig:continuum}).

{\em \cii\ line:} In order to measure the size of the \cii\ emitting
region, we first created a continuum subtracted data cube: we fitted a
first order polynomial to the channels at least 0.75\,GHz away from
the line centre and subtracted this continuum from the data using the
CASA task ``uvcontsub". A map of the \cii\ emission was produced by
averaging the cube over
600\,\kms\ (1.5\,$\times$\,FWHM$_\mathrm{[CII]}$) around the peak of
the \cii\ emission. The result is shown in
Figure~\ref{fig:ciimaps}. From a 2D Gaussian fit to this map we
obtained a deconvolved size of the \cii\ emitting region of
(0.31$\pm$0.05)$\times$(0.27$\pm$0.05)\,arcsec$^2$ in diameter, which

\begin{deluxetable*}{lcccc}
\tablecaption{Derived Properties of the Host Galaxy of Quasar J1120+0641 at $z=7.0851$ as Function of Temperature $T_d$ and Emissivity Index $\beta$ \label{tab:prop}}
\tablehead{
\colhead{} & \colhead{$T_d=47$\,K, $\beta=1.6$, no CMB} & \colhead{$T_d=47$\,K, $\beta=1.6$} & \colhead{$T_d=41$\,K, $\beta=1.95$} & \colhead{$T_d=30$\,K, $\beta=1.6$}
}
\startdata
\lfir [\lsun] & $(1.3\pm0.1)\times10^{12}$ & $(1.5\pm0.1)\times10^{12}$ & $(1.5\pm0.1)\times10^{12}$ & $(5.6\pm0.4)\times10^{11}$ \\
\ltir [\lsun] & $(1.9\pm0.1)\times10^{12}$ & $(2.1\pm0.2)\times10^{12}$ & $(1.9\pm0.1)\times10^{12}$ & $(7.7\pm0.6)\times10^{11}$ \\ 
$M_\mathrm{dust}$ [\msun] & $(8.6\pm0.6)\times10^7~$ & $(9.6\pm0.7)\times10^7~$ & $(7.7\pm0.6)\times10^7~$ & $(4.2\pm0.3)\times10^8~$ \\
\lcii [\lsun] & $(1.3\pm0.1)\times10^9~$ & $(1.5\pm0.1)\times10^9~$ & $(1.6\pm0.1)\times10^9~$ & $(2.0\pm0.2)\times10^9~$ \\
\lci [\lsun] & $<$$1.0\times10^8$ & $<$$1.4\times10^8$ & $<$1.5$\times10^8$ & $<$$2.3\times10^8$ \\
$L_\mathrm{CO(2-1)}$ [\lsun] & $<$$5.0\times10^6$ & $<$$8.4\times10^6$ & $<$$9.4\times10^6$ & $<$$1.6\times10^7$ \\
$L_\mathrm{CO(7-6)}$ [\lsun] & $<$$1.0\times10^8$ & $<$$1.4\times10^8$ & $<$$1.5\times10^8$ & $<$$2.3\times10^8$ \\
\lcii/\lci & $>$13.0 & $>$10.6 & $>$10.1 & $>$8.7 \\
$L^\prime_\mathrm{CO(1-0)}$\tablenotemark{a} [K\,\kms\,pc$^2$] & $<$$1.3\times10^{10}$ & $<$$2.1\times10^{10}$ & $<$$2.4\times10^{10}$ & $<$$4.0\times10^{10}$ \\
\lcii/$L_\mathrm{CO(1-0)}$\tablenotemark{a} & $>$2200 & $>$1400 & $>$1300 & $>$1000 \\
$M_\mathrm{gas}/M_\mathrm{dust}$ & $<$120 & $<$180 & $<$250 & $<$80 \\
SFR$_\mathrm{TIR}$ [\msunyr] & 280$\pm$20 & 315$\pm$25 & 290$\pm$20 & 115$\pm$10 \\
SFR$_\mathrm{[CII]}$ [\msunyr] & 70--440 & 80--500 & 85--525 & 110--700 \\
\enddata
\tablenotetext{a}{Derived from the 3\,$\sigma$ limit on the CO(2-1) emission and assuming the molecular gas is thermalized ($L^\prime_\mathrm{CO(2-1)}=L^\prime_\mathrm{CO(1-0)}$). Taking instead the limit on the CO(7-6) emission and adopting a CO spectral line energy distribution similar to that observed in the quasar J1148+5251 at $z=6.42$ gives upper limits on the CO(1-0) emission that are a factor 1.0--1.4 smaller.}
\end{deluxetable*}

\noindent
corresponds to (1.65$\pm$0.29)$\times$(1.44$\pm$0.26)\,kpc$^2$. The
area of the resolved \cii\ emitting region of 1.9$\pm$0.5\,kpc$^2$ is
larger than the area of the continuum region (0.8$\pm$0.2\,kpc$^2$). A
larger extent of \cii\ emission compared to that of the continuum
emission has also been reported in other $z\gtrsim6$ quasar host
galaxies \citep[e.g.,][]{wan13,ven16}. The total flux density of the
resolved component is $1.35\pm0.15$\,mJy, which corresponds to a flux
of $0.81\pm0.09$\,Jy\,\kms.

We also performed aperture photometry on the \cii\ image. We recovered
all the flux measured in the tapered spectrum within a radius of
0\farcs8 ($\sim$4.3\,kpc). At larger aperture radii we tentatively
detected additional flux. We estimated that up to 20\% of additional
line flux might be present at scales up to $\sim$7\,kpc from the
quasar, although the significance is low (1$\sigma$).

Bright \cii\ emission in $z\gtrsim6$ quasar hosts often shows
indications of rotation \citep[e.g.,][]{wil13,wan13,ven16}. To
investigate whether the gas in J1120+0641 displays ordered motion, we
separately mapped the blue and red side of the emission line, see
Figure~\ref{fig:ciimaps}. The blue- and redshifted emission peaks
coincide with the continuum emission. It therefore appears that the
gas traced by \cii\ emission does not show rotation on scales of
$\gtrsim$1\,kpc. We will discuss the implications of this in
Section~\ref{sec:mdyn}.

A summary of our results, described in
Sections~\ref{sec:fircontinuum}--\ref{sec:size}, is listed in
Table~\ref{tab:res}.

\section{DISCUSSION}
\label{sec:discussion}

\subsection{Origin of the Heating Radiation}
\label{sec:pdr}

We now compare our (limits on the) emission line ratios to models to
constrain the physical parameters of the emitting gas \citep[see,
  e.g.,][]{kau99,mei05,mei07}. In particular, the line ratio
\cii/\ci\ can be used to determine the dominant source of radiation,
the hard X-ray radiation of the accreting, supermassive black hole (an
X-ray dominated region, or XDR) or UV radiation from hot stars (a
photon dominated region, or PDR) (see
Figure~\ref{fig:ciicilineratio}). To calculate the intrinsic line
ratios, we first need to determine the intrinsic luminosity of the
emission lines. Because we are measuring the flux of the lines against
the CMB, the intrinsic luminosity will depend on the excitation
temperature \citep{dac13}: $F_\mathrm{in} = F_\mathrm{obs} /
\{1-B_{\nu}[T_\mathrm{CMB}(z=7.0851)]/B_{\nu}[T_\mathrm{ex}]\}$, where
$F_\mathrm{obs}$ and $F_\mathrm{in}$ are the observed and intrinsic
line flux, $B_\nu$ the Planck function at the rest-frame frequency
$\nu$ of the line, $T_\mathrm{CMB}(z=7.0851)$ the temperature of the
CMB at redshift $z=7.0851$ ($\approx22$\,K) and $T_\mathrm{ex}$ the
excitation temperature.

If collisions dominate the excitation, then the excitation temperature
is set by the kinetic temperature of the gas. In this paper we further
assume thermodynamic equilibrium between the dust and the gas, i.e.,
$T_\mathrm{gas}=T_\mathrm{dust}$. This assumption is motivated by the
study of dust and CO emission in the host galaxy of quasar J1148+5251
at $z=6.42$ in which $T_\mathrm{ex} \approx T_\mathrm{dust}$
\citep{bee06,rie09,stef15}. We also assume that the dust has a
constant temperature throughout the host galaxy.  To explore the range
of luminosities and line ratios in J1120+0641, we derive intrinsic
luminosities of the emission lines for various temperatures that are
found in the literature (Table~\ref{tab:prop}). Several studies of
$z\sim6$ quasar host galaxies \citep[e.g.,][]{wil13,wil15,wan13}
implement a dust temperature of $T_d=47$\,K as derived by
\citet{bee06} for distant luminous quasars. A study by \citet{pri01}
found an average dust temperature in quasars of $T_d=41$\,K. We also
computed the line luminosities in the case of a lower temperature of
$T_d=30$\,K \citep{wal11,ven16}. Alternatively, the gas temperature
could be much higher than that of the dust, $T_\mathrm{gas}\gg100$\,K
\citep[e.g.,][]{con13}, and the effect of the CMB become negligible
(the ``no CMB" column in Table~\ref{tab:prop}).

\begin{figure*}
\figurenum{4}
\plotone{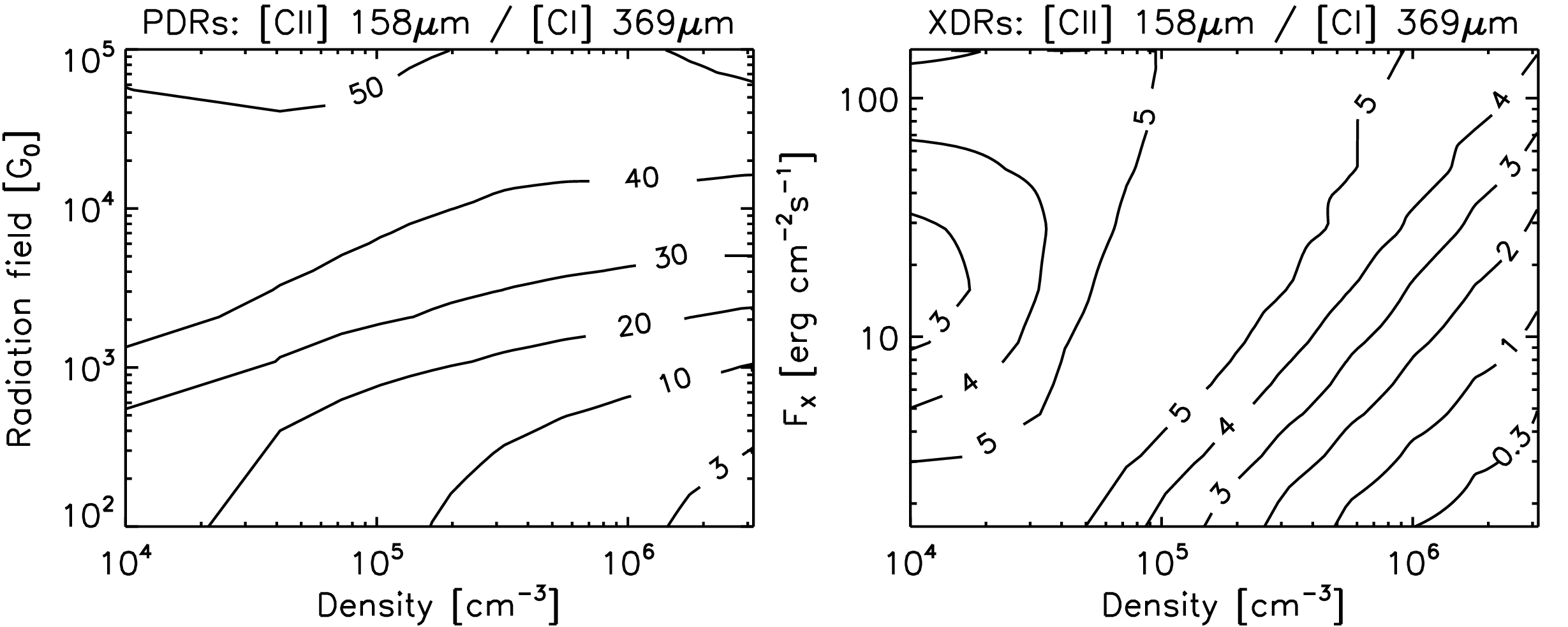}
\caption{{\em Left:} \cii\ over \ci\ line ratio as function of density
  and radiation field in case of a photon dominated region (PDR),
  adapted from \citet{mei05,mei07}. {\em Right:} Same line ratio, but
  this time for an X-ray dominated region (XDR), also adapted from
  \citet{mei05,mei07}. The maximum ratio expected in an XDR is
  $\sim$6, independent of (realistic) radiation strength and
  density. According to these models, our measured lower limit of
  $\sim$8.7 (Table~\ref{tab:prop}) excludes that the X-ray radiation
  of the quasar dominates the gas heating.}
\label{fig:ciicilineratio}
\end{figure*}

We can compare the limits on the \cii/\ci\ ratio in J1120+0641
(Table~\ref{tab:prop}) to those from PDR and XDR models
(Figure~\ref{fig:ciicilineratio}). In PDR models, the \cii/\ci\ ratio
covers a large range from $\sim$3--50, depending on the density and
strength of the UV radiation field \citep{kau99,mei07}. On the other
hand, in XDRs the \cii/\ci\ ratio is generally lower than in a PDR,
with a maximum of around $\sim$6 \citep{mei07}. We measure a lower
limit on the \cii/\ci\ line ratio $\sim$8.7. Therefore, based on the
XDR models, we can exclude that the radiation illuminating the gas is
dominated by hard X-ray radiation from the accreting black hole, but
instead should mainly come UV from stars.

\subsection{Infrared Luminosity and Star-Formation Rates}
\label{sec:sfrd}

To compute the far-infrared luminosity, we have to assume a shape of
the dust emission. The cold dust spectral energy distribution (SED) of
distant quasar host galaxies is often parameterized as an optically
thin modified black body \citep[e.g.,][]{pri01,bee06,lei14} with a
dust temperature $T_d$ and emissivity index $\beta$. Adopting
$T_d=47$\,K and $\beta=1.6$ \citep{bee06}, integrating the dust SED
from 42.5\,$\mu$m to 122.5\,$\mu$m, and taking the CMB into account,
we derive a FIR luminosity of
\lfir\,$=(1.5\pm0.1)\times10^{12}$\,\lsun. For $T_d=41$\,K and
$\beta=1.95$ \citep{pri01}, we derive a similar value for the FIR
luminosity.  A lower dust temperature of 30\,K \citep{ven16} and
$\beta=1.6$ results in \lfir\,$=(5.6\pm0.4)\times10^{11}$\,\lsun\ (see
Table~\ref{tab:prop} for a summary).

In Section~\ref{sec:pdr} we concluded that the gas is predominantly
heated by UV radiation from stars. If that also applies to the dust,
then we can use the infrared luminosity to constrain the
star-formation rate (SFR) of the host galaxy. This is supported by
\citet{bar15} who analyzed the full SED of J1120+0641 and concluded
that the emission around 235\,GHz in the rest-frame is dominated by a
cool dust component and not by the accreting black hole.

To estimate the SFR from the continuum detection, we first integrated
the modified black body from 8\,$\mu$m to 1000\,$\mu$m to obtain the
total infrared luminosity \ltir. Depending on the parameters of the
modified black body, we derive total infrared luminosities ranging
from \lfir\,$=(7.7\pm0.6)\times10^{11}$\,\lsun\ to
\lfir\,$=(2.1\pm0.2)\times10^{12}$\,\lsun\ (Table~\ref{tab:prop}). We
then applied the local scaling relation between the total infrared
luminosity and SFR from \citet{mur11}: SFR$_\mathrm{TIR} =
L_\mathrm{TIR} / 6.7\times10^{9}$ with SFR$_\mathrm{TIR}$ in units of
\msunyr\ and \ltir\ in units of \lsun. We estimate a SFR of
105--340\,\msunyr\ (Table~\ref{tab:prop}), where the main uncertainty
is the shape of the dust SED. In Section~\ref{sec:size} we concluded
that $\sim$80\% of the continuum emission originates from a region
that measures 1.2\,kpc\,$\times$\,0.8\,kpc in diameter, or
0.8\,kpc$^2$. This means that the star-formation rate density (SFRD)
is $\sim$100--350\,\msunyr\,kpc$^2$. The lower limit is an order of
magnitude smaller than the SFRD derived for the more FIR--luminous
bright quasar J1148+5251 at $z=6.42$ \citep{wal09b}, in which the SFRD
approaches the Eddington limit for star formation.

Alternatively, we can compute the SFR from the luminosity of the
\cii\ line. Applying the \cii\ SFR conversion for high redshift
sources from \citet{del14}, we derive a
SFR$_\mathrm{[CII]}=70-700$\,\msunyr\ (Table~\ref{tab:prop}). Within
the large uncertainties, this value is consistent with the one derived
from the FIR continuum emission.

\subsection{Dust and Gas Mass}
\label{sec:dustgas}

Following \citet{ven12} we derived the dust mass from the FIR
luminosity assuming a temperature and a dust mass opacity coefficient:
$M_\mathrm{dust}\sim S_\nu / [\kappa_\lambda \times B_\nu(T_d)]$
\citep[e.g.,][]{hil83} with $S_\nu$ the continuum flux density at
rest-frame frequency $\nu$ and the dust mass opacity $\kappa_\lambda =
0.77 (850\,\mu\mathrm{m}/\lambda)^\beta$\,cm$^2$\,g$^{-1}$
\citep{dun00}. For the range of temperatures and emissivity indices
considered in this paper, our best estimate for the dust mass in
J1120+0641 is $(0.8-4)\times10^8$\,\msun\ (Table~\ref{tab:prop}).

A limit on the molecular gas mass can be derived from the upper limit
on the CO luminosity $L^\prime_\mathrm{CO(2-1)}$ (in units of
K\,\kms\,pc$^2$) using $L^\prime_\mathrm{CO(1-0)}\approx
L^\prime_\mathrm{CO(2-1)}$ \citep[e.g.,][]{car13} and applying a
conversion factor of
$\alpha=M_\mathrm{gas}/L^\prime_\mathrm{CO(1-0)}=0.8$\,\msun\,(K\,\kms\,pc$^2$)$^{-1}$
as found for nearby ULIRGs \citep[e.g.,][]{dow98}. Taking the effects
of the CMB into account, our VLA upper limit on the CO(2-1) line flux
results in an upper limit on the CO luminosity of
$L^\prime_\mathrm{CO(1-0)}<4\times10^{10}$\,K\,\kms\,pc$^2$ (see
Table~\ref{tab:prop}) and a gas mass limit of
$M_\mathrm{gas}<3\times10^{10}$\,\msun. Alternatively, we can use the
PdBI upper limit on the CO(7-6) line flux to derive
$L^\prime_\mathrm{CO(1-0)}$. Applying an CO excitation ladder that is
similar to that in quasar J1146+5251 at $z=6.42$
\citep[e.g.,][]{stef15}, we derive an upper limit of
$M_\mathrm{gas}\lesssim2\times10^{10}$\,\msun.

If we combine these upper limits with our estimates of the dust mass,
we derive gas-to-dust mass ratio limits of $<$80--250 (strongly
depending on the dust and gas temperature, see Table~\ref{tab:prop}),
which are consistent with locally derived values of $\sim$100
\citep[e.g.,][]{dra07,san13}.

\subsection{Dynamical Mass Estimate}
\label{sec:mdyn}

The dynamical mass of $z>6$ quasar hosts has often been computed by
assuming that the gas is rotating in an inclined disk
\citep[e.g.,][]{wal03,wan13,wil15,ven16}. This approach was motivated
by the detection of velocity gradients in the \cii\ emitting gas. From
Figure~\ref{fig:ciimaps} it is clear that in J1120+0641 there is no
evidence for rotation on scales of $\gtrsim$1\,kpc. Instead, we here
use the virial theorem to estimate a dynamical mass of the quasar host
galaxy: $M_{\mathrm{dyn}}=3R\sigma^2 / 2G$, with $R$ the radius of the
line emitting region, $\sigma$ the velocity dispersion of the gas and
$G$ the gravitational constant. In Section~\ref{sec:size} we measure a
maximum radius of 4.3\,kpc and from the tapered spectrum in
Figure~\ref{fig:spectrum} we obtain a velocity dispersion of
$\sigma=169\pm18$\,\kms, which results in an upper limit on the
dynamical mass of $M_\mathrm{dyn}=(4.3\pm0.9)\times10^{10}$\,\msun,
which is similar to dynamical masses derived for other $z\gtrsim6$
quasar host galaxies
\citep[e.g.,][]{wal09b,wan13,wil15,ven16}. Intriguingly, the black
hole with a mass of
$M_\mathrm{BH}=(2.4\pm0.2)\times10^9$\,\msun\footnote{The uncertainty
  quoted here does not include the systematic uncertainty of a factor
  $\sim$3 in the method applied by \citet{der14} to derive the mass of
  black hole.} \citep{der14} already contains $\sim$6\% of this
dynamical mass. This fraction is about 10$\times$ higher than the mass
ratio of black holes and bulges in local early type galaxies
\citep[\mbh/$M_\mathrm{bulge}=0.49$\%,][]{kor13}. High black hole to
dynamical mass ratios are also found in other $z\gtrsim6$ quasar host
galaxies, which have black hole to dynamical mass ratios up to
$M_\mathrm{BH}/M_\mathrm{dyn} \lesssim 25$\% and an average of
$\langle M_\mathrm{BH}/M_\mathrm{dyn} \rangle \approx 2$\%
\citep[][and references therein]{ven16}. The high ratios disagree with
some simulations of high redshift quasar host galaxies
\citep[e.g.,][]{val14}. Various solutions for this discrepancy have
been proposed in the literature. As nearly all high redshift quasars
are selected from flux limited, optical/near-infrared surveys, by
design only the most massive black holes with the highest accretion
rates are selected. Also, due to scatter in the correlation between
black hole and galaxy mass, the massive black holes of $z\gtrsim6$
quasars are preferentially found in galaxies that can be an order of
magnitude less massive than expected based on the correlation itself
\citep[e.g.,][]{wil05b}. Alternatively, far-infrared emission lines
only trace the gas in the inner regions of a galaxy and using these
lines could underestimate the true dynamical mass of the quasar host
\citep[e.g.,][]{val14}.

We can also estimate the dynamical mass of the central, unresolved
emission. The gas within the central beam with a FWHM of 0\farcs23
(1.2\,kpc) has a velocity dispersion of
$\sigma=235\pm25$\,\kms. Setting the radius of this region to
$R=0.5\times$\,FWHM$_\mathrm{beam}$, we derive a mass of
$M_\mathrm{dyn,centre}=(1.2\pm0.2)\times10^{10}$\,\msun, which is only
$\sim$5 times more than the mass of the black hole.

We can compare the dynamical mass with the mass of the molecular gas
in the quasar host galaxy. Assuming a gas-to-dust mass ratio of 100,
the gas mass is $(0.8-4)\times10^{10}$\,\msun, which is 20--95\% of
the dynamical mass. In Section~\ref{sec:size} we showed that a high
fraction ($\sim$80\%) of the dust and \cii\ emission resides in a
compact region with a diameter $<$1.5\,kpc. If this is also the case
for the molecular gas, then there is not much room for a massive
stellar component in the central $\sim$1--1.5\,kpc of the host galaxy,
which raises interesting questions regarding the origin of the
detected dust emission. Due to the large uncertainties in the
molecular gas mass, the black hole mass and dynamical mass, we cannot
put any strong constraints on the stellar mass in the quasar host
galaxy.

\section{Summary}
\label{sec:summary}

We present ALMA, PdBI, and VLA observations targeting the dust
emission and \cii, \ci, and two CO lines in the host galaxy of quasar
J1120+0641 at $z=7.1$. The ALMA observations of the \cii\ line and the
underlying continuum greatly increase the spatial resolution compared
to earlier measurements with the PdBI (factor of 70 in beam area).
Our main findings can be summarized as follows.

\begin{itemize}

\item Within the field of view mapped by ALMA ($\sim$25\arcsec) the
  quasar is the only detected source.

\item The dust continuum and \cii\ emission regions are very compact
  and only marginally resolved in the ALMA data. The majority of the
  emission (80\%) is associated with a compact region of size
  1.2$\times$0.8\,kpc$^2$ in diameter.

\item The non--detection of the \ci\ line indicates that the heating
  in the quasar host galaxy is dominated by star formation (PDR), and
  not by the accreting supermassive black hole (XDR).

\item From the limits on the CO(2-1) and CO(7-6) lines we derived
  upper limits on the molecular gas mass of
  $M_\mathrm{gas}<4\times10^{10}$\,\msun. This is consistent with a
  gas-to-dust mass ratio around $\sim$100 that is measured in the
  local Universe.

\item We estimate the star formation rate in the quasar host using
  both the FIR continuum and the \cii\ line measurement. Both methods
  give consistent results (SFR$_\mathrm{FIR}\sim$105--340\,\msunyr)
  with a resulting star formation rate surface density of
  $\sim$100--350\,\msunyr\,kpc$^{-2}$, well below the value for
  Eddington--accretion--limited star formation \citep{wal09b}

\item Surprisingly, the compact \cii\ emission does not exhibit
  ordered motion on kpc-scales: applying the virial theorem yields a
  dynamical mass of the host galaxy of
  $(4.3\pm0.9)\times10^{10}$\,\msun, only $\sim$20$\times$ higher than
  that of the central supermassive black hole. In the very central
  region, the dynamical mass of the host is only 5 times that of the
  central black hole. In this region, the mass of the black hole and
  that of the implied dust and gas is able to explain the dynamical
  mass. In other words, there is not much room for a massive stellar
  component in the very central region. However, we note that the
  large uncertainties in both the molecular gas and dust mass, and the
  dynamical mass make it unfeasable to put tight constraints on the
  stellar mass.

\end{itemize}

The ALMA observations presented here start to spatially resolve the
host galaxy of the most distant quasar known. With the recent
commissioning of even longer baselines at ALMA, even higher resolution
imaging of this quasar host is now possible that will start to
spatially resolve the sphere of influence of the central supermassive
black hole.

\acknowledgments 

We thank the referee for providing valuable comments and
suggestions. B.P.V. and F.W. acknowledge funding through the ERC grant
``Cosmic Dawn''. Support for R.D.  was provided by the DFG priority
program 1573 ``The physics of the interstellar medium''.  PCH and RGM
acknowledge the support of the UK Science and Technology Facilities
Council (STFC) through the award of a Consolidated Grant to the
Institute of Astronomy. We thank Rowin Meijerink for help with the
modeling of the line ratios. This paper makes use of the following
ALMA data: ADS/JAO.ALMA\#2012.1.00882.S. ALMA is a partnership of ESO
(representing its member states), NSF (USA) and NINS (Japan), together
with NRC (Canada) and NSC and ASIAA (Taiwan), in cooperation with the
Republic of Chile. The Joint ALMA Observatory is operated by ESO,
AUI/NRAO and NAOJ. Based in part on observations carried out with the
IRAM Plateau de Bure Interferometer. IRAM is supported by INSU/CNRS
(France), MPG (Germany), and IGN (Spain). The National Radio Astronomy
Observatory is a facility of the National Science Foundation operated
under cooperative agreement by Associated Universities, Inc.

\facilities{ALMA, IRAM:Interferometer, EVLA}

\listofchanges


\begin{thebibliography}{}
\expandafter\ifx\csname natexlab\endcsname\relax\def\natexlab#1{#1}\fi

\bibitem[{{Aravena} {et~al.}(2016){Aravena}, {Decarli}, {Walter}, {Da Cunha},
  {Bauer}, {Carilli}, {Daddi}, {Elbaz}, {Ivison}, {Riechers}, {Smail},
  {Swinbank}, {Weiss}, {Anguita}, {Assef}, {Bell}, {Bertoldi}, {Bacon},
  {Bouwens}, {Cortes}, {Cox}, {G{\'o}nzalez-L{\'o}pez}, {Hodge}, {Ibar},
  {Inami}, {Infante}, {Karim}, {Le Le F{\`e}vre}, {Magnelli}, {Ota}, {Popping},
  {Sheth}, {van der Werf}, \& {Wagg}}]{ara16}
{Aravena}, M., {Decarli}, R., {Walter}, F., {et~al.} 2016, ApJ, 833, 68

\bibitem[{{Ba{\~n}ados} {et~al.}(2016){Ba{\~n}ados}, {Venemans}, {Decarli},
  {Farina}, {Mazzucchelli}, {Walter}, {Fan}, {Stern}, {Schlafly}, {Chambers},
  {Rix}, {Jiang}, {McGreer}, {Simcoe}, {Wang}, {Yang}, {Morganson}, {De Rosa},
  {Greiner}, {Balokovi{\'c}}, {Burgett}, {Cooper}, {Draper}, {Flewelling},
  {Hodapp}, {Jun}, {Kaiser}, {Kudritzki}, {Magnier}, {Metcalfe}, {Miller},
  {Schindler}, {Tonry}, {Wainscoat}, {Waters}, \& {Yang}}]{ban16}
{Ba{\~n}ados}, E., {Venemans}, B.~P., {Decarli}, R., {et~al.} 2016, ApJS, 227,
  11

\bibitem[{{Barnett} {et~al.}(2015){Barnett}, {Warren}, {Banerji}, {McMahon},
  {Hewett}, {Mortlock}, {Simpson}, {Venemans}, {Ota}, \& {Shibuya}}]{bar15}
{Barnett}, R., {Warren}, S.~J., {Banerji}, M., {et~al.} 2015, A\&A, 575, A31

\bibitem[{{Beelen} {et~al.}(2006){Beelen}, {Cox}, {Benford}, {Dowell},
  {Kov{\'a}cs}, {Bertoldi}, {Omont}, \& {Carilli}}]{bee06}
{Beelen}, A., {Cox}, P., {Benford}, D.~J., {et~al.} 2006, ApJ, 642, 694

\bibitem[{{Bertoldi} {et~al.}(2003){Bertoldi}, {Carilli}, {Cox}, {Fan},
  {Strauss}, {Beelen}, {Omont}, \& {Zylka}}]{ber03a}
{Bertoldi}, F., {Carilli}, C.~L., {Cox}, P., {et~al.} 2003, A\&A, 406, L55

\bibitem[{{Carilli} \& {Holdaway}(1999)}]{car99}
{Carilli}, C.~L., \& {Holdaway}, M.~A. 1999, Radio Science, 34, 817

\bibitem[{{Carilli} \& {Walter}(2013)}]{car13}
{Carilli}, C.~L., \& {Walter}, F. 2013, ARA\&A, 51, 105

\bibitem[{{Carnall} {et~al.}(2015){Carnall}, {Shanks}, {Chehade}, {Fumagalli},
  {Rauch}, {Irwin}, {Gonzalez-Solares}, {Findlay}, \& {Metcalfe}}]{carn15}
{Carnall}, A.~C., {Shanks}, T., {Chehade}, B., {et~al.} 2015, MNRAS, 451, L16

\bibitem[{{Contursi} {et~al.}(2013){Contursi}, {Poglitsch}, {Gr{\'a}cia
  Carpio}, {Veilleux}, {Sturm}, {Fischer}, {Verma}, {Hailey-Dunsheath}, {Lutz},
  {Davies}, {Gonz{\'a}lez-Alfonso}, {Sternberg}, {Genzel}, \&
  {Tacconi}}]{con13}
{Contursi}, A., {Poglitsch}, A., {Gr{\'a}cia Carpio}, J., {et~al.} 2013, A\&A,
  549, A118

\bibitem[{{da Cunha} {et~al.}(2013){da Cunha}, {Groves}, {Walter}, {Decarli},
  {Weiss}, {Bertoldi}, {Carilli}, {Daddi}, {Elbaz}, {Ivison}, {Maiolino},
  {Riechers}, {Rix}, {Sargent}, \& {Smail}}]{dac13}
{da Cunha}, E., {Groves}, B., {Walter}, F., {et~al.} 2013, ApJ, 766, 13

\bibitem[{{De Looze} {et~al.}(2014){De Looze}, {Cormier}, {Lebouteiller},
  {Madden}, {Baes}, {Bendo}, {Boquien}, {Boselli}, {Clements}, {Cortese},
  {Cooray}, {Galametz}, {Galliano}, {Graci{\'a}-Carpio}, {Isaak}, {Karczewski},
  {Parkin}, {Pellegrini}, {R{\'e}my-Ruyer}, {Spinoglio}, {Smith}, \&
  {Sturm}}]{del14}
{De Looze}, I., {Cormier}, D., {Lebouteiller}, V., {et~al.} 2014, A\&A, 568,
  A62

\bibitem[{{De Rosa} {et~al.}(2011){De Rosa}, {Decarli}, {Walter}, {Fan},
  {Jiang}, {Kurk}, {Pasquali}, \& {Rix}}]{der11}
{De Rosa}, G., {Decarli}, R., {Walter}, F., {et~al.} 2011, ApJ, 739, 56

\bibitem[{{De Rosa} {et~al.}(2014){De Rosa}, {Venemans}, {Decarli}, {Gennaro},
  {Simcoe}, {Dietrich}, {Peterson}, {Walter}, {Frank}, {McMahon}, {Hewett},
  {Mortlock}, \& {Simpson}}]{der14}
{De Rosa}, G., {Venemans}, B.~P., {Decarli}, R., {et~al.} 2014, ApJ, 790, 145

\bibitem[{{D{\'{\i}}az-Santos} {et~al.}(2013){D{\'{\i}}az-Santos}, {Armus},
  {Charmandaris}, {Stierwalt}, {Murphy}, {Haan}, {Inami}, {Malhotra},
  {Meijerink}, {Stacey}, {Petric}, {Evans}, {Veilleux}, {van der Werf}, {Lord},
  {Lu}, {Howell}, {Appleton}, {Mazzarella}, {Surace}, {Xu}, {Schulz},
  {Sanders}, {Bridge}, {Chan}, {Frayer}, {Iwasawa}, {Melbourne}, \&
  {Sturm}}]{dia13}
{D{\'{\i}}az-Santos}, T., {Armus}, L., {Charmandaris}, V., {et~al.} 2013, ApJ,
  774, 68

\bibitem[{{Downes} \& {Solomon}(1998)}]{dow98}
{Downes}, D., \& {Solomon}, P.~M. 1998, ApJ, 507, 615

\bibitem[{{Draine} {et~al.}(2007){Draine}, {Dale}, {Bendo}, {Gordon}, {Smith},
  {Armus}, {Engelbracht}, {Helou}, {Kennicutt}, {Li}, {Roussel}, {Walter},
  {Calzetti}, {Moustakas}, {Murphy}, {Rieke}, {Bot}, {Hollenbach}, {Sheth}, \&
  {Teplitz}}]{dra07}
{Draine}, B.~T., {Dale}, D.~A., {Bendo}, G., {et~al.} 2007, ApJ, 663, 866

\bibitem[{{Dunne} {et~al.}(2000){Dunne}, {Eales}, {Edmunds}, {Ivison},
  {Alexander}, \& {Clements}}]{dun00}
{Dunne}, L., {Eales}, S., {Edmunds}, M., {et~al.} 2000, MNRAS, 315, 115

\bibitem[{{Fan} {et~al.}(2006){Fan}, {Strauss}, {Becker}, {White}, {Gunn},
  {Knapp}, {Richards}, {Schneider}, {Brinkmann}, \& {Fukugita}}]{fan06b}
{Fan}, X., {Strauss}, M.~A., {Becker}, R.~H., {et~al.} 2006, AJ, 132, 117

\bibitem[{{Hildebrand}(1983)}]{hil83}
{Hildebrand}, R.~H. 1983, QJRAS, 24, 267

\bibitem[{{Jiang} {et~al.}(2007){Jiang}, {Fan}, {Vestergaard}, {Kurk},
  {Walter}, {Kelly}, \& {Strauss}}]{jia07}
{Jiang}, L., {Fan}, X., {Vestergaard}, M., {et~al.} 2007, AJ, 134, 1150

\bibitem[{{Jiang} {et~al.}(2015){Jiang}, {McGreer}, {Fan}, {Bian}, {Cai},
  {Cl{\'e}ment}, {Wang}, \& {Fan}}]{jia15}
{Jiang}, L., {McGreer}, I.~D., {Fan}, X., {et~al.} 2015, AJ, 149, 188

\bibitem[{{Jiang} {et~al.}(2009){Jiang}, {Fan}, {Bian}, {Annis}, {Chiu},
  {Jester}, {Lin}, {Lupton}, {Richards}, {Strauss}, {Malanushenko},
  {Malanushenko}, \& {Schneider}}]{jia09}
{Jiang}, L., {Fan}, X., {Bian}, F., {et~al.} 2009, AJ, 138, 305

\bibitem[{{Kaufman} {et~al.}(1999){Kaufman}, {Wolfire}, {Hollenbach}, \&
  {Luhman}}]{kau99}
{Kaufman}, M.~J., {Wolfire}, M.~G., {Hollenbach}, D.~J., \& {Luhman}, M.~L.
  1999, ApJ, 527, 795

\bibitem[{{Kormendy} \& {Ho}(2013)}]{kor13}
{Kormendy}, J., \& {Ho}, L.~C. 2013, ARA\&A, 51, 511

\bibitem[{{Kroupa}(2001)}]{kro01}
{Kroupa}, P. 2001, MNRAS, 322, 231

\bibitem[{{Kurk} {et~al.}(2007){Kurk}, {Walter}, {Fan}, {Jiang}, {Riechers},
  {Rix}, {Pentericci}, {Strauss}, {Carilli}, \& {Wagner}}]{kur07}
{Kurk}, J.~D., {Walter}, F., {Fan}, X., {et~al.} 2007, ApJ, 669, 32

\bibitem[{{Leipski} {et~al.}(2014){Leipski}, {Meisenheimer}, {Walter}, {Klaas},
  {Dannerbauer}, {De Rosa}, {Fan}, {Haas}, {Krause}, \& {Rix}}]{lei14}
{Leipski}, C., {Meisenheimer}, K., {Walter}, F., {et~al.} 2014, ApJ, 785, 154

\bibitem[{{Maiolino} {et~al.}(2005){Maiolino}, {Cox}, {Caselli}, {Beelen},
  {Bertoldi}, {Carilli}, {Kaufman}, {Menten}, {Nagao}, {Omont}, {Wei{\ss}},
  {Walmsley}, \& {Walter}}]{mai05}
{Maiolino}, R., {Cox}, P., {Caselli}, P., {et~al.} 2005, A\&A, 440, L51

\bibitem[{{Matsuoka} {et~al.}(2016){Matsuoka}, {Onoue}, {Kashikawa}, {Iwasawa},
  {Strauss}, {Nagao}, {Imanishi}, {Niida}, {Toba}, {Akiyama}, {Asami}, {Bosch},
  {Foucaud}, {Furusawa}, {Goto}, {Gunn}, {Harikane}, {Ikeda}, {Kawaguchi},
  {Kikuta}, {Komiyama}, {Lupton}, {Minezaki}, {Miyazaki}, {Morokuma},
  {Murayama}, {Nishizawa}, {Ono}, {Ouchi}, {Price}, {Sameshima}, {Silverman},
  {Sugiyama}, {Tait}, {Takada}, {Takata}, {Tanaka}, {Tang}, \&
  {Utsumi}}]{mat16}
{Matsuoka}, Y., {Onoue}, M., {Kashikawa}, N., {et~al.} 2016, ApJ, 828, 26

\bibitem[{{McMullin} {et~al.}(2007){McMullin}, {Waters}, {Schiebel}, {Young},
  \& {Golap}}]{mul07}
{McMullin}, J.~P., {Waters}, B., {Schiebel}, D., {Young}, W., \& {Golap}, K.
  2007, in Astronomical Society of the Pacific Conference Series, Vol. 376,
  Astronomical Data Analysis Software and Systems XVI, ed. R.~A. {Shaw},
  F.~{Hill}, \& D.~J. {Bell}, 127

\bibitem[{{Meijerink} \& {Spaans}(2005)}]{mei05}
{Meijerink}, R., \& {Spaans}, M. 2005, A\&A, 436, 397

\bibitem[{{Meijerink} {et~al.}(2007){Meijerink}, {Spaans}, \& {Israel}}]{mei07}
{Meijerink}, R., {Spaans}, M., \& {Israel}, F.~P. 2007, A\&A, 461, 793

\bibitem[{{Momjian} {et~al.}(2014){Momjian}, {Carilli}, {Walter}, \& {Venemans}}]{mom14}
{Momjian}, E., {Carilli}, C.~L., {Walter}, F., \& {Venemans}, B. 2014, AJ, 147, 6

\bibitem[{{Mortlock} {et~al.}(2009){Mortlock}, {Patel}, {Warren}, {Venemans},
  {McMahon}, {Hewett}, {Simpson}, {Sharp}, {Burningham}, {Dye}, {Ellis},
  {Gonzales-Solares}, \& {Hu{\'e}lamo}}]{mor09}
{Mortlock}, D.~J., {Patel}, M., {Warren}, S.~J., {et~al.} 2009, A\&A, 505, 97

\bibitem[{{Mortlock} {et~al.}(2011){Mortlock}, {Warren}, {Venemans}, {Patel},
  {Hewett}, {McMahon}, {Simpson}, {Theuns}, {Gonz{\'a}les-Solares}, {Adamson},
  {Dye}, {Hambly}, {Hirst}, {Irwin}, {Kuiper}, {Lawrence}, \&
  {R{\"o}ttgering}}]{mor11}
{Mortlock}, D.~J., {Warren}, S.~J., {Venemans}, B.~P., {et~al.} 2011, Nature,
  474, 616

\bibitem[{{Murphy} {et~al.}(2011){Murphy}, {Condon}, {Schinnerer}, {Kennicutt},
  {Calzetti}, {Armus}, {Helou}, {Turner}, {Aniano}, {Beir{\~a}o}, {Bolatto},
  {Brandl}, {Croxall}, {Dale}, {Donovan Meyer}, {Draine}, {Engelbracht},
  {Hunt}, {Hao}, {Koda}, {Roussel}, {Skibba}, \& {Smith}}]{mur11}
{Murphy}, E.~J., {Condon}, J.~J., {Schinnerer}, E., {et~al.} 2011, ApJ, 737, 67

\bibitem[{{Priddey} \& {McMahon}(2001)}]{pri01}
{Priddey}, R.~S., \& {McMahon}, R.~G. 2001, MNRAS, 324, L17

\bibitem[{{Reed} {et~al.}(2015){Reed}, {McMahon}, {Banerji}, {Becker},
  {Gonzalez-Solares}, {Martini}, {Ostrovski}, {Rauch}, {Abbott}, {Abdalla},
  {Allam}, {Benoit-Levy}, {Bertin}, {Buckley-Geer}, {Burke}, {Carnero Rosell},
  {da Costa}, {D'Andrea}, {DePoy}, {Desai}, {Diehl}, {Doel}, {Cunha},
  {Estrada}, {Evrard}, {Fausti Neto}, {Finley}, {Fosalba}, {Frieman}, {Gruen},
  {Honscheid}, {James}, {Kent}, {Kuehn}, {Kuropatkin}, {Lahav}, {Maia},
  {Makler}, {Marshall}, {Merritt}, {Miquel}, {Mohr}, {Nord}, {Ogando},
  {Plazas}, {Romer}, {Roodman}, {Rykoff}, {Sako}, {Sanchez}, {Santiago},
  {Schubnell}, {Sevilla}, {Smith}, {Soares-Santos}, {Suchyta}, {Swanson},
  {Tarle}, {Thomas}, {Tucker}, {Walker}, \& {Wechsler}}]{ree15}
{Reed}, S.~L., {McMahon}, R.~G., {Banerji}, M., {et~al.} 2015, MNRAS, 454, 3952

\bibitem[{{Riechers} {et~al.}(2009){Riechers}, {Walter}, {Bertoldi}, {Carilli},
  {Aravena}, {Neri}, {Cox}, {Wei{\ss}}, \& {Menten}}]{rie09}
{Riechers}, D.~A., {Walter}, F., {Bertoldi}, F., {et~al.} 2009, ApJ, 703, 1338

\bibitem[{{Rujopakarn} {et~al.}(2016){Rujopakarn}, {Dunlop}, {Rieke}, {Ivison},
  {Cibinel}, {Nyland}, {Jagannathan}, {Silverman}, {Alexander}, {Biggs},
  {Bhatnagar}, {Ballantyne}, {Dickinson}, {Elbaz}, {Geach}, {Hayward},
  {Kirkpatrick}, {McLure}, {Micha{\l}owski}, {Miller}, {Narayanan}, {Owen},
  {Pannella}, {Papovich}, {Pope}, {Rau}, {Robertson}, {Scott}, {Swinbank}, {van
  der Werf}, {van Kampen}, {Weiner}, \& {Windhorst}}]{ruj16}
{Rujopakarn}, W., {Dunlop}, J.~S., {Rieke}, G.~H., {et~al.} 2016, ApJ, 833, 12

\bibitem[{{Sandstrom} {et~al.}(2013){Sandstrom}, {Leroy}, {Walter}, {Bolatto},
  {Croxall}, {Draine}, {Wilson}, {Wolfire}, {Calzetti}, {Kennicutt}, {Aniano},
  {Donovan Meyer}, {Usero}, {Bigiel}, {Brinks}, {de Blok}, {Crocker}, {Dale},
  {Engelbracht}, {Galametz}, {Groves}, {Hunt}, {Koda}, {Kreckel}, {Linz},
  {Meidt}, {Pellegrini}, {Rix}, {Roussel}, {Schinnerer}, {Schruba}, {Schuster},
  {Skibba}, {van der Laan}, {Appleton}, {Armus}, {Brandl}, {Gordon}, {Hinz},
  {Krause}, {Montiel}, {Sauvage}, {Schmiedeke}, {Smith}, \& {Vigroux}}]{san13}
{Sandstrom}, K.~M., {Leroy}, A.~K., {Walter}, F., {et~al.} 2013, ApJ, 777, 5

\bibitem[{{Sargsyan} {et~al.}(2014){Sargsyan}, {Samsonyan}, {Lebouteiller},
  {Weedman}, {Barry}, {Bernard-Salas}, {Houck}, \& {Spoon}}]{sar14}
{Sargsyan}, L., {Samsonyan}, A., {Lebouteiller}, V., {et~al.} 2014, ApJ, 790,
  15

\bibitem[{{Stefan} {et~al.}(2015){Stefan}, {Carilli}, {Wagg}, {Walter},
  {Riechers}, {Bertoldi}, {Green}, {Fan}, {Menten}, \& {Wang}}]{stef15}
{Stefan}, I.~I., {Carilli}, C.~L., {Wagg}, J., {et~al.} 2015, MNRAS, 451, 1713

\bibitem[{{Valiante} {et~al.}(2014){Valiante}, {Schneider}, {Salvadori}, \& {Gallerani}}]{val14}
{Valiante}, R., {Schneider}, R., {Salvadori}, S., \& {Gallerani}, S. 2014, MNRAS, 444, 2442

\bibitem[{{Venemans} {et~al.}(2016){Venemans}, {Walter}, {Zschaechner},
  {Decarli}, {De Rosa}, {Findlay}, {McMahon}, \& {Sutherland}}]{ven16}
{Venemans}, B.~P., {Walter}, F., {Zschaechner}, L., {et~al.} 2016, ApJ, 816, 37

\bibitem[{{Venemans} {et~al.}(2012){Venemans}, {McMahon}, {Walter}, {Decarli},
  {Cox}, {Neri}, {Hewett}, {Mortlock}, {Simpson}, \& {Warren}}]{ven12}
{Venemans}, B.~P., {McMahon}, R.~G., {Walter}, F., {et~al.} 2012, ApJL, 751,
  L25

\bibitem[{{Venemans} {et~al.}(2013){Venemans}, {Findlay}, {Sutherland}, {De
  Rosa}, {McMahon}, {Simcoe}, {Gonz{\'a}lez-Solares}, {Kuijken}, \&
  {Lewis}}]{ven13}
{Venemans}, B.~P., {Findlay}, J.~R., {Sutherland}, W.~J., {et~al.} 2013, ApJ,
  779, 24

\bibitem[{{Venemans} {et~al.}(2015){Venemans}, {Ba{\~n}ados}, {Decarli},
  {Farina}, {Walter}, {Chambers}, {Fan}, {Rix}, {Schlafly}, {McMahon},
  {Simcoe}, {Stern}, {Burgett}, {Draper}, {Flewelling}, {Hodapp}, {Kaiser},
  {Magnier}, {Metcalfe}, {Morgan}, {Price}, {Tonry}, {Waters}, {AlSayyad},
  {Banerji}, {Chen}, {Gonz{\'a}lez-Solares}, {Greiner}, {Mazzucchelli},
  {McGreer}, {Miller}, {Reed}, \& {Sullivan}}]{ven15a}
{Venemans}, B.~P., {Ba{\~n}ados}, E., {Decarli}, R., {et~al.} 2015, ApJL, 801,
  L11

\bibitem[{{Walter} {et~al.}(2009){Walter}, {Riechers}, {Cox}, {Neri},
  {Carilli}, {Bertoldi}, {Weiss}, \& {Maiolino}}]{wal09b}
{Walter}, F., {Riechers}, D., {Cox}, P., {et~al.} 2009, Nature, 457, 699

\bibitem[{{Walter} {et~al.}(2011){Walter}, {Wei{\ss}}, {Downes}, {Decarli}, \&
  {Henkel}}]{wal11}
{Walter}, F., {Wei{\ss}}, A., {Downes}, D., {Decarli}, R., \& {Henkel}, C.
  2011, ApJ, 730, 18

\bibitem[{{Walter} {et~al.}(2003){Walter}, {Bertoldi}, {Carilli}, {Cox}, {Lo},
  {Neri}, {Fan}, {Omont}, {Strauss}, \& {Menten}}]{wal03}
{Walter}, F., {Bertoldi}, F., {Carilli}, C., {et~al.} 2003, Nature, 424, 406

\bibitem[{{Wang} {et~al.}(2008){Wang}, {Carilli}, {Wagg}, {Bertoldi}, {Walter},
  {Menten}, {Omont}, {Cox}, {Strauss}, {Fan}, {Jiang}, \& {Schneider}}]{wan08b}
{Wang}, R., {Carilli}, C.~L., {Wagg}, J., {et~al.} 2008, ApJ, 687, 848

\bibitem[{{Wang} {et~al.}(2013){Wang}, {Wagg}, {Carilli}, {Walter}, {Lentati},
  {Fan}, {Riechers}, {Bertoldi}, {Narayanan}, {Strauss}, {Cox}, {Omont},
  {Menten}, {Knudsen}, {Neri}, \& {Jiang}}]{wan13}
{Wang}, R., {Wagg}, J., {Carilli}, C.~L., {et~al.} 2013, ApJ, 773, 44

\bibitem[{{Willott} {et~al.}(2015){Willott}, {Bergeron}, \& {Omont}}]{wil15}
{Willott}, C.~J., {Bergeron}, J., \& {Omont}, A. 2015, ApJ, 801, 123

\bibitem[{{Willott} {et~al.}(2013){Willott}, {Omont}, \& {Bergeron}}]{wil13}
{Willott}, C.~J., {Omont}, A., \& {Bergeron}, J. 2013, ApJ, 770, 13

\bibitem[{{Willott} {et~al.}(2010{\natexlab{a}}){Willott}, {Albert},
  {Arzoumanian}, {Bergeron}, {Crampton}, {Delorme}, {Hutchings}, {Omont},
  {Reyl{\'e}}, \& {Schade}}]{wil10b}
{Willott}, C.~J., {Albert}, L., {Arzoumanian}, D., {et~al.} 2010{\natexlab{a}},
  AJ, 140, 546

\bibitem[{{Willott} {et~al.}(2010{\natexlab{b}}){Willott}, {Delorme},
  {Reyl{\'e}}, {Albert}, {Bergeron}, {Crampton}, {Delfosse}, {Forveille},
  {Hutchings}, {McLure}, {Omont}, \& {Schade}}]{wil10a}
{Willott}, C.~J., {Delorme}, P., {Reyl{\'e}}, C., {et~al.} 2010{\natexlab{b}},
  AJ, 139, 906

\bibitem[{{Willott} {et~al.}(2005){Willott}, {Percival}, {McLure}, {Crampton}, {Hutchings}, {Jarvis}, {Sawicki}, \& {Simard}}]{wil05b}
{Willott}, C.~J., {Percival}, W.~J., {McLure}, R.~J., {et~al.} 2005, ApJ, 626, 657

\bibitem[{{Wu} {et~al.}(2015){Wu}, {Wang}, {Fan}, {Yi}, {Zuo}, {Bian}, {Jiang},
  {McGreer}, {Wang}, {Yang}, {Yang}, {Thompson}, \& {Beletsky}}]{wu15}
{Wu}, X.-B., {Wang}, F., {Fan}, X., {et~al.} 2015, \nat, 518, 512

\end{thebibliography}
\end{document}